\documentclass[preprint]{aastex}

\renewcommand{\cite}{\citep}

\newcommand{\tohoku}{\affil{Research Center for Neutrino Science, Tohoku University, Sendai 980-8578, Japan}}

\newcommand{\alabama}{\affil{Department of Physics and Astronomy, University of Alabama, Tuscaloosa, Alabama 35487, USA}}

\newcommand{\lbl}{\affil{Physics Department, University of California, Berkeley and \\
Lawrence Berkeley National Laboratory,  Berkeley, California 94720, USA}}

\newcommand{\caltech}{\affil{W.~K.~Kellogg Radiation Laboratory, California Institute of Technology, Pasadena, California 91125, USA}}
 
\newcommand{\colostate}{\affil{Department of Physics, Colorado State University, Fort Collins, Colorado 80523, USA}}

\newcommand{\drexel}{\affil{Physics Department, Drexel University, Philadelphia, Pennsylvania 19104, USA}}

\newcommand{\hawaii}{\affil{Department of Physics and Astronomy, University of Hawaii at Manoa, Honolulu, Hawaii 96822, USA}}

\newcommand{\kansas}{\affil{Department of Physics, Kansas State University, Manhattan, Kansas 66506, USA}}

\newcommand{\stanford}{\affil{Physics Department, Stanford University, Stanford, California 94305, USA}}

\newcommand{\ut}{\affil{Department of Physics and Astronomy, University of Tennessee, Knoxville, Tennessee 37996, USA}}

\newcommand{\tunl}{\affil{Triangle Universities Nuclear Laboratory, Durham, North Carolina 27708, USA and \\
Physics Departments at Duke University, North Carolina Central University, and the University of North Carolina at Chapel Hill}}

\newcommand{\wisc}{\affil{Department of Physics, University of Wisconsin, Madison, Wisconsin 53706, USA}}

\newcommand{\ipmu}{\affil{Institute for the Physics and Mathematics of the Universe, University of Tokyo, Kashiwa 277-8568, Japan}}

\newcommand{\nikhef}{\affil{Nikhef, Science Park, Amsterdam, the Netherlands}}

\shortauthors{Gando et al.}

\begin{document}

\title{Search for extraterrestrial antineutrino sources with the KamLAND detector}

% Tohoku

\author{
A.~Gando,
Y.~Gando,
K.~Ichimura,
H.~Ikeda,
K.~Inoue\altaffilmark{1},
Y.~Kibe\altaffilmark{2},
Y.~Kishimoto,
M.~Koga\altaffilmark{1},
Y.~Minekawa,
T.~Mitsui,
T.~Morikawa,
N.~Nagai,
K.~Nakajima
K.~Nakamura\altaffilmark{1},
K.~Narita,
I.~Shimizu,
Y.~Shimizu,
J.~Shirai,
F.~Suekane,
A.~Suzuki,
H.~Takahashi,
N.~Takahashi,
Y.~Takemoto,
K.~Tamae,
H.~Watanabe,
B.D.~Xu,
H.~Yabumoto,
H.~Yoshida,
and S.~Yoshida
}\tohoku

% IPMU

\author{
S.~Enomoto\altaffilmark{3},
A.~Kozlov,
and H.~Murayama\altaffilmark{4}
}\ipmu

% Alabama

\author{
C.~Grant,
G.~Keefer\altaffilmark{5},
and A.~Piepke\altaffilmark{1}
}\alabama

% LBNL and UC Berkeley

\author{
T.I.~Banks,
T.~Bloxham,
J.A.~Detwiler,
S.J.~Freedman\altaffilmark{1},
B.K.~Fujikawa\altaffilmark{1},
K.~Han,
R.~Kadel,
T.~O'Donnell,
and H.M.~Steiner
}\lbl

% Caltech

\author{
D.A.~Dwyer,
R.D.~McKeown,
and C.~Zhang
}\caltech

% Colorado State

\author{
B.E.~Berger
}\colostate

% Drexel

\author{
C.E.~Lane,
J.~Maricic,
and T.~Miletic\altaffilmark{6},
}\drexel

% Hawaii

\author{
M.~Batygov\altaffilmark{7},
J.G.~Learned,
S.~Matsuno,
and M.~Sakai
}\hawaii

% KSU

\author{
G.A.~Horton-Smith\altaffilmark{1}
}\kansas

% Stanford
\author{
K.E.~Downum and
G.~Gratta
}\stanford

% UT

\author{
Y.~Efremenko\altaffilmark{1},
Y.~Kamyshkov,
and O.~Perevozchikov\altaffilmark{8},
}\ut

% TUNL

\author{
H.J.~Karwowski,
D.M.~Markoff,
and W.~Tornow
}\tunl

% Wisconsin

\author{
K.M.~Heeger\altaffilmark{1} 
}\wisc

\and

% Nikhef

\author{
M.P.~Decowski\altaffilmark{1}
}\nikhef

\altaffiltext{1}{Also with the Institute for the Physics and Mathematics of the Universe, University of Tokyo, Kashiwa 277-8568, Japan}

\altaffiltext{2}{Present address: Department of Physics, Tokyo Institute of Technology, Tokyo 152-8551, Japan}

\altaffiltext{3}{Also with the Center for Experimental Nuclear Physics and Astrophysics, University of Washington, Seattle, Washington 98195, USA}

\altaffiltext{4}{Also with the Physics Department, University of California, Berkeley and Lawrence Berkeley National Laboratory,  Berkeley, California 94720, USA}

\altaffiltext{5}{Present address: Lawrence Livermore National Laboratory, Livermore, California 94550, USA}

\altaffiltext{6}{Present address: Department of Physics and Astronomy, Rowan University, 201 Mullica Hill Road, Glassboro, New Jersey 08028, USA}

\altaffiltext{7}{Present address: SNOLAB, Lively, ON P3Y 1M3, Canada}

\altaffiltext{8}{Present address: Louisiana State University Department of Physics \& Astronomy
Nicholson Hall, Tower Dr. Baton Rouge, LA 70803-4001, USA}

\begin{abstract}
We present the results of a search for extraterrestrial electron antineutrinos ($\overline{\nu}_{e}$'s) in the energy range \mbox{$8.3\,{\rm MeV} < E_{\overline{\nu}_{e}} < 31.8\,{\rm MeV}$} using the KamLAND detector.
In an exposure of 4.53\,kton-year, we
identify 25 candidate events. All of the candidate events can be attributed to background,
most importantly neutral current atmospheric neutrino interactions, 
setting an upper limit on the probability of $^{8}{\rm B}$ solar $\nu_{e}$'s converting into $\overline{\nu}_{e}$'s at \mbox{$5.3\times10^{-5}$}\ (90\% CL), if we assume an undistorted $\overline{\nu}_{e}$ shape.
This limit corresponds to a solar $\overline{\nu}_{e}$ flux of $93\,{\rm cm^{-2} s^{-1}}$ or an event rate of $1.6\ {\rm events\,(kton\mathchar`-year)^{-1}}$ above the energy threshold \mbox{$(E_{\overline{\nu}_{e}} \ge 8.3\,{\rm MeV})$}.
The present data also allows us to set more stringent limits on the diffuse supernova neutrino flux and on the annihilation rates for light dark matter particles. 
\end{abstract}

\keywords{dark matter, ISM: supernova remnants, neutrinos, Sun: particle emission}

\section{Introduction}

Ambient electron antineutrinos ($\overline{\nu}_{e}$'s) of terrestrial
origin include geoneutrinos~\cite{Araki2005b, Bellini2010}, which have energies
below $\sim$3.4\,MeV, and
man-made reactor antineutrinos~\cite{Cowan1956,Vogel1981}, which have energies below $\sim$8\,MeV.
Naturally produced $\overline{\nu}_{e}$'s with higher energies must be
of cosmic origin.
The region above a few tens of MeV\ is dominated by atmospheric neutrinos
generated from the decays of muons and pions produced by cosmic-ray interactions.
There is a gap in energy between terrestrial and atmospheric neutrinos where other ``exotic'' mechanisms to generate neutrinos could dominate.
For antineutrinos in the energy region between 8 MeV and 15\,MeV, 
only diffuse neutrino flux from distant supernovae~\cite{Totani1995} and exotic generation mechanisms, e.g., conversion of solar neutrinos into antineutrinos~\cite{Okun1986} 
or light dark matter annihilation~\cite{Palomares2008} are thought to be possible.

\subsection{Solar Antineutrinos}

The nuclear fusion reaction produces most of the Sun's energy, and a portion of its energy is taken away by electron neutrinos. The data for all solar neutrino experiments are consistent with the prediction based on the Standard Solar Model including the flavor transition due to the neutrino oscillation.
On the other hand, antineutrinos are produced by the $\beta^{-}$ decay of the natural radioactivity or the photo-fission of heavy isotopes in the Sun, but both flux contributions at Earth's surface are negligibly small relative to the terrestrial antineutrinos~\cite{Malaney1990}.

However, if the neutrino has a non-zero magnetic moment it could be converted into an antineutrino in the strong solar magnetic field. 
This mechanism was originally proposed as a solution to the solar
neutrino problem~\cite{Okun1986} and was later revisited in~\cite{Akhmedov2003}.
A two-step process takes place,
the first step occurs deep inside the solar interior, where a $\nu_{e}$ converts into a $\overline{\nu}_{\mu}$ via spin flavor precession.
The $\overline{\nu}_{\mu}$ then oscillates into an
$\overline{\nu}_{e}$ while propagating from the Sun to the Earth.
The combined probability for the two processes is
\begin{equation}
\label{equation:FlavorPrecession}
P(\nu_{eL}\rightarrow\overline{\nu}_{eR}) \simeq
1.8\times10^{-10} \sin^2 2 \theta_{12}
\left[
\frac{\mu}{10^{-12}\mu_{B}}
\frac{B_{T}(0.05R_{\odot})}{10\:\mbox{kG}}
\right]^{2},
\end{equation}
where $B_{T}$\ is the transverse solar magnetic field in the region of neutrino production, $R_{\odot}$ is the solar radius, and $\mu$\ is the neutrino magnetic moment in Bohr magneton ($\mu_{B}$).
Very little is known about the magnitude of magnetic fields in the solar interior,
but values up to $3\times10^{7}$\,G are permitted based on {\it Solar and Heliospheric Observatory} observations~\cite{Couvidat2003}.
One can search for conversion of $^{8}{\rm B}$ solar neutrinos because they have energies higher than terrestrial neutrinos.
The present best limit for the probability of solar neutrino-to-antineutrino
conversion, from the Borexino experiment, is less than $1.3\times10^{-4}$~\cite{Bellini2011} 
assuming an unoscillated $^{8}{\rm B}$ neutrino flux of $5.88 \times 10^{6}\,{\rm cm^{-2} s^{-1}}$~\cite{Serenelli2010}.
The neutrino decay, predicting that a heavy neutrino mass eigenstate decays to a lighter antineutrino mass eigenstate, is another possibility of antineutrino production~\cite{Beacom2002}.

\subsection{Diffuse Supernova Neutrino Flux}

A diffuse supernova neutrino background exists from distant core-collapse supernovae. 
The numerical calculation of supernova explosions predicts neutrino fluxes which are comparable to each other, 
and those energy spectra extend up to $\sim$80 MeV.
In typical hydrogen-rich detectors, the supernova neutrino signal is dominated by the $\overline{\nu}_{e}$ reaction due to the large cross section. 
Only upper limits for the diffuse supernova neutrino flux (DSNF) have
been set by Super-Kamiokande and the Sudbury Neutrino Observatory (SNO) for energies above 19.3\,MeV~\cite{Malek2003} and 22.9\,MeV~\cite{Aharmim2006} for $\overline{\nu}_{e}$ and ${\nu}_{e}$, respectively.
Because the DSNF energy is lowered by redshift, the spectral shape is closely connected to the history of star formation.
Various supernova and cosmological models predict different shapes~\cite{Ando2004}, while recent measurements of the core-collapse supernova rate reduce the cosmological uncertainties~\cite{Horiuchi2009, Beacom2010}. The model predictions  should be confronted with data.

\subsection{Dark Matter Annihilation}

Dark matter self-annihilation into standard model particles can be searched experimentally.
In light dark matter annihilation, most of the final state particles are kinematically forbidden, and only $\gamma\gamma$, $e^{+}e^{-}$, and $\overline{\nu}\nu$ are possible. However, there are strong limits on the branching ratio to the visible states, $\gamma\gamma$ and $e^{+}e^{-}$~\cite{Mack2008}, so the $\overline{\nu}\nu$ search bounds the total annihilation cross section. 
Because even a few MeV mass cold dark matter particles are non-relativistic, their annihilation should produce monoenergetic fluxes of neutrinos. Annihilation in the Milky Way halo should be the dominant source of such neutrinos, so redshift can be safely neglected.
The Super-Kamiokande data provide the best limit on the rate of such annihilation for dark matter mass ($m_{\chi}$) above 15\,MeV. 
Assuming an annihilation model in the Galactic
halo and equal annihilation into all neutrino flavors~\cite{Palomares2008}, the limit can be translated into a
velocity-dependent averaged cross section ($\left< \sigma_{A} v
\right>$) for dark matter particles.

\section{The KamLAND Experiment}
\label{section:KamlandExperiment}

\begin{figure}
\begin{center}
\includegraphics[angle=0,width=\columnwidth]{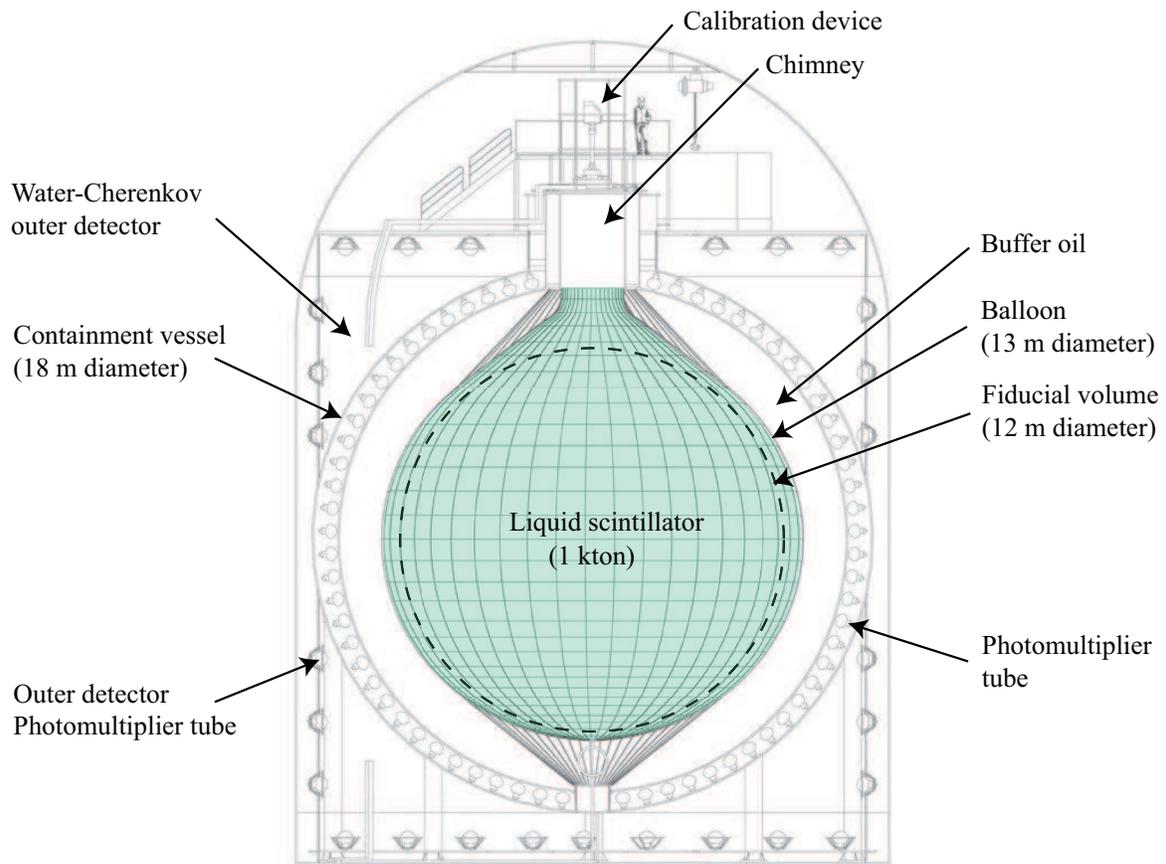}
\end{center}
\caption{Schematic diagram of the KamLAND detector.}
\label{figure:KamLAND}
\end{figure}

The KamLAND detector is located $\sim$1\,km under the peak of Mt. Ikenoyama
($36.42^{\circ}$N, $137.31^{\circ}$E) near Kamioka, Japan.
The 2700 meters water equivalent (mwe) of vertical rock overburden reduces the cosmic-ray muon flux by almost five orders of magnitude.
A schematic diagram of KamLAND is shown in Figure~\ref{figure:KamLAND}.
The primary target volume consists of 1\,kton of ultra-pure liquid scintillator~(LS) contained in a 13\,m diameter spherical balloon made of
135\,$\mu$m thick transparent nylon EVOH (ethylene vinyl alcohol copolymer) composite film.
The LS consists of 80\% dodecane and
20\% pseudocumene (1,2,4-trimethylbenzene) by volume,
and \mbox{$1.36\pm0.03$}\,${\rm g}\ {\rm l}^{-1}$ of the fluor PPO (2,5-diphenyloxazole). 
A buffer comprising 57\% isoparaffin and 43\% dodecane oils by volume, which fills the region between the balloon
and the surrounding 18\,m diameter spherical stainless-steel outer vessel, shields the LS from external radiation.
The specific gravity of the buffer oil is adjusted to be 0.04\% lower than that of the LS.
An array of photomultiplier tubes (PMTs)---1325 specially developed fast PMTs masked to 17\,inch diameter and
554 older 20\,inch diameter PMTs reused from the Kamiokande experiment~\cite{Kume1983}---are mounted on the inner surface of the stainless steel,
providing 34\% photocathode coverage.
This inner detector is shielded by a 3.2\,kton water-Cherenkov outer detector (OD). 

Electron antineutrinos are detected in KamLAND via the inverse beta-decay reaction, 
\begin{equation}
\mbox{$\overline{\nu}_{e}+p\rightarrow e^{+}+n$}.
\end{equation}
This process has a delayed-coincidence~(DC) event-pair signature which offers powerful background suppression. 
The energy deposited by the positron, which generates the DC pair's
prompt event, is approximately related to the incident
$\overline{\nu}_{e}$ energy by \mbox{$E_{\overline{\nu}_{e}} \simeq E_{\mathrm{p}} + \overline{E}_{n} + 0.8\,\mbox{MeV}$}, where $E_{\mathrm{p}} (\equiv T_{e^{+}} + 2 m_{e})$ is the sum of the $e^{+}$ kinetic energy and
annihilation $\gamma$ energies, and $\overline{E}_{n}$ is the average
neutron recoil energy which is below 1 MeV
for the analysis $E_{\rm p}$ range, 7.5--30.0\,MeV.
Most of the neutron recoil energy is transferred to recoil protons, resulting in a negligible contribution of scintillation light due to the quenching effect.
The delayed event in the DC pair is generated by a 2.2\,MeV $\gamma$-ray produced when the neutron captures on a proton.
The mean neutron capture time is $(207.5 \pm 2.8)\,\mu$s~\cite{Abe2010}\,.
The inverse beta decay cross section is well approximated at the first order in $1/M$~\cite{Vogel1999}, where $M$ is the nucleon mass. The angular distribution of the positron emission is nearly isotropic, and unlike water Cherenkov detector, the scintillation light is also isotropic.
As a result, the positron signal does not provide the incoming antineutrino source direction. Due to the extremely low cross section of antineutrinos, the Earth does not make a shadow extraterrestrial antineutrinos, and the detector has isotropic sensitivity.

The detector is periodically calibrated with $\gamma$ sources deployed from a glove box installed at the top of the chimney region.
The radioactive sources are $^{60}$Co, $^{68}$Ge, $^{203}$Hg,
$^{65}$Zn, $^{241}$Am$^{9}$Be, $^{137}$Cs, and $^{210}$Po$^{13}$C,
providing energy calibration up to $\sim$8\,MeV along the central axis
of the detector.
In addition, capture of neutrons on hydrogen and carbon provides energy calibration throughout the entire sensitive volume.
The visible energy in the detector is measured from the number of detected photoelectrons and is corrected for event position,
detector non-uniformity, and scintillator nonlinearity from quenching and Cherenkov light production.
The overall vertex reconstruction resolution is $\sim$\mbox{12\,cm~$/\sqrt{E(\mbox{MeV})}$}
and energy resolution is \mbox{$6.4\%/ \sqrt{E(\mbox{MeV})}$}.
Energy reconstruction of positrons with $E_{\rm p} > 7.5$\,MeV (i.e., $E_{\overline{\nu}_e} > 8.3$\,MeV) is verified using tagged $^{12}$B $\beta^{-}$-decays (\mbox{$\tau=29.1$\,ms} and \mbox{$Q=13.4$\,MeV}) generated via muon
spallation~\cite{Abe2010}. 

From the studies of Bi-Po sequential decays, the effective equilibrium concentration of $^{238}$U and $^{232}$Th in the LS are $(2.2 \pm 0.3) \times 10^{-18}\,{\rm g}\ {\rm g}^{-1}$ and $(4.8 \pm 0.3) \times 10^{-17}\,{\rm g}\ {\rm g}^{-1}$, prior to the start of the LS purification campaign in 2007. From the singles energy spectrum fit, the concentration of $^{40}$K is $(2.2 \pm 0.2) \times 10^{-16}\,{\rm g}\ {\rm g}^{-1}$. Those radioactive impurities are negligible in this study relative to other backgrounds, such as muon spallation products and external $\gamma$-rays. The LS purification further reduced the radioactive impurities.

\section{Event Selection}
\label{section:EventSelection}

The present analysis includes data accumulated between 2002 March 5 and 2010 July 23, corresponding to 2343 live-days.
For the present search the following criteria were used: the prompt energy is required to be \mbox{$7.5\,{\rm MeV}<E_{\rm p}<30.0\,{\rm MeV}$}, and the delayed energy to be \mbox{$1.8\,{\rm MeV}<E_{\rm d}<2.6\,{\rm MeV}$}; a fiducial volume cut of \mbox{$R<6$\,m} on both prompt and delayed events, a time correlation cut of \mbox{$0.5\,\mu{\rm s} <\Delta T<1000\,\mu{\rm s}$}, and a spatial correlation cut of \mbox{$\Delta R<1.6$\,m}, this cut is driven by the mean free path of the capture-$\gamma$ in the LS rather than the diffusion distance of the neutron.
The inefficiency caused by neutrons escaping from the LS is negligibly small.
We limited the prompt energy window up to 30 MeV, because in the higher energy, there are background events from oil Cherenkov muons which are untagged by the OD due to a small inefficiency.
Spallation cuts were used to reduce backgrounds from long-lived isotopes,
such as $^{9}$Li (\mbox{$\tau=257$\,ms} and \mbox{$Q=13.6$\,MeV}), that are generated by cosmic muons passing through the scintillator:
a 2\,ms veto is applied to the entire detector volume after a non-showering muon for both prompt and delayed events,
a 2\,s veto is applied after a showering muon (i.e., muons depositing more than 3\,GeV of energy above their minimum ionizing contribution) or non-reconstructed muon, while
a 2\,s 3\,m radius cylindrical cut is applied around well-reconstructed non-showering muons~\cite{Abe2010} for delayed events.
The overall selection efficiency of the candidates is 92\%, which is evaluated from a Monte Carlo (MC) simulation.

\section{Background Calculations}

\subsection{Random Coincidences}

Two uncorrelated events in the detector may accidentally coincide in time, space, and energy so as to pass the $\overline{\nu}_{e}$ selection cuts.
To estimate the background contribution from random coincidences,
events were selected with the appropriate prompt and delayed energies but in an out-of-time interval of 0.2\,s to 1.2\,s after the prompt event.
This out-of-time window is $10^{3}$\ times longer than the time interval used for the $\overline{\nu}_{e}$ selection,
providing a high-statistics background measurement.
The time distribution between prompt and delayed events in the range between 0.2\,s and 1.2\,s shows no correlation between these events.
The random coincidence background for the analysis is determined to be \mbox{$0.22\pm0.01$}\ DC-pairs.

\subsection{Reactor Antineutrinos}

The location of the KamLAND detector was selected for the copious $\overline{\nu}_{e}$ flux from 56 Japanese nuclear power plants in order to study neutrino
oscillation~\cite{Gando2011}.
The reactor $\overline{\nu}_{e}$ flux at KamLAND dominates all other $\overline{\nu}_{e}$ sources for $E_{\rm p}\,<\,7.5$\,MeV.
However, the tail of the reactor neutrino energy distribution extends to higher energies.
The $\overline{\nu}_{e}$ flux comes primarily from the beta decay of neutron-rich fragments produced in the fission of four isotopes: $^{235}$U,
$^{238}$U,
$^{239}$Pu,
and $^{241}$Pu.
For each reactor the appropriate operational records
including thermal power generation, fuel burn-up, shutdowns and fuel reload schedule were used to calculate the fission rates.
The resulting $\overline{\nu}_{e}$ spectrum was calculated using the model of~\cite{Schreckenbach1985, Hahn1989, Vogel1981} taking neutrino oscillation into account.
The same methodology was used for previous reactor $\overline{\nu}_{e}$ analyses and
showed excellent agreement over a wide energy range between expected and detected $\overline{\nu}_{e}$ events~\cite{Gando2011}.
The total number of reactor $\overline{\nu}_{e}$ candidates having
$E_{\rm p} > 7.5$\,MeV is calculated to be \mbox{$2.2\pm0.7$}\ events, including a $\sim$10\% event-rate increase due to energy resolution.

\subsection{Radioactive Isotopes}

Cosmic-ray muons interacting with carbon nuclei in the scintillator produce a variety of radioactive isotopes~\cite{Abe2010}.
Two of these isotopes,
$^{8}$He (\mbox{$\tau=171.7$\,ms} and \mbox{$Q=10.7$\,MeV}) and 
$^{9}$Li (\mbox{$\tau=257.2$\,ms} and \mbox{$Q=13.6$\,MeV}),
have decay modes with electrons and neutrons in the final state.
Such decays create DC pairs similar to
inverse beta decay and therefore represent a background in the present study.

The combination of a 2\,s veto of the detector after showering muons and
a 2\,s  3\,m radius cylindrical cut after non-showering muons
significantly reduces the contribution of this background, but cannot eliminate it completely.
The $^{9}$Li isotope, which has a higher end-point value, longer life time and a higher production rate, generates the majority of these background events after cuts.
To determine the contribution from this background we selected $^{9}$Li candidates using the same cuts that were used for the selection of the $\overline{\nu}_{e}$ candidates,
but the muon veto was not applied.
The $^{9}$Li rate was evaluated from the distribution of the decay time
relative to all previous muons, using a wider energy window of $0.9$\,MeV$ <
E_{\rm p} < 15.0$\,MeV to reduce statistical errors. 
For a 6\,m fiducial volume, \mbox{$2074\pm49$}\ events were found after showering muons and \mbox{$454\pm31$}\ events in a 3\,m radius cylinder around the muon track after non-showering muons. Twenty percent of these events occur in the energy region of interest $7.5$\,MeV$ < E_{\rm p} < 15.0$\,MeV.
The 2\,s cut reduces the $^{9}$Li background from showering muons to less than 0.2\ events for $E_{\rm p} > 7.5$\,MeV.
As non-showering muons occur with a relatively high frequency (0.2 Hz), to avoid the drastic loss of exposure which would accompany a 2\,s full-detector veto we take advantage of the fact that non-showering muons can be relatively well tracked in the LS and instead restrict the 2\,s veto to a 3\,m radius cylinder around the muon track.
A 2\,ms full-volume veto after all tagged muons is also used to suppress any spallation neutrons.
To measure the efficiency of these cuts the distribution of neutron captures as a function of distance from the muon track was examined, and we found only 5.9\% of neutrons survive the 3\,m radius cylindrical cut. 
The resulting number of $^{9}$Li from spallation background with $E_{\rm p} > 7.5$\,MeV and surviving the 3\,m radius cylindrical cut and time cut is \mbox{$4.0\pm0.3$}\ events. The dead time introduced by all these cuts is 9.2\%.

\subsection{Fast Neutrons}

Fast neutrons outside the inner detector may cause backgrounds in the fiducial volume. A fast neutron can scatter on protons or carbon nuclei in the LS producing a scintillation signal followed by a neutron capture signal, mimicking an $\overline{\nu}_{e}$ coincidence. A MC simulation of fast neutrons reveals that the dominant background contribution is caused by muon-induced cosmogenic neutrons. A 2\,ms veto after OD-tagged muons mostly eliminates this background, while OD-untagged muons and the OD inefficiency cause a residual background. The MC-based study estimates $3.2 \pm 3.2$ fast neutrons remain in the data set, where a conservative uncertainty of 100\% for the simulated neutron production rate by muons is assumed~\cite{Abe2010}.

\subsection{Atmospheric Neutrino Interactions}

Charged current (CC) and neutral current (NC) interactions of atmospheric neutrinos with carbon atoms in the KamLAND scintillator are the most significant source of background.
Atmospheric neutrino spectra from~\cite{Honda2007},
calculated specifically for the KamLAND location,
were used to estimate the contribution from these backgrounds.

The CC reactions by atmospheric $\overline{\nu}_{e}$'s generate an irreducible background.
The contribution from atmospheric $\overline{\nu}_{e}$'s is estimated to be $\sim$0.06\ events in the energy window $7.5$\,MeV$ < E_{\rm p} < 30.0$\,MeV, which is dominated by the reaction on protons, because the cross section for carbon nuclei is estimated to be at least one order of magnitude smaller~\cite{Kim2009}.
Atmospheric $\nu_{\mu}$'s and $\overline{\nu}_{\mu}$'s could react with both protons and carbon nuclei to produce muons and neutrons.
The amount of detectable energy is shifted lower for such reactions because a large fraction of the neutrino's initial energy is expended to produce the muon.
On the other hand, such reactions are followed by muon decay and therefore manifest themselves as a triple time correlation between the prompt event,
muon decay and neutron capture.
In the event selection, we found one triple coincidence event accompanied by a muon decay signal 
in a decay time interval of 0.5--10\,$\mu{\rm s}$. 
This event was excluded from the candidates.
To calculate the contribution from these reactions,
the cross sections from~\cite{Athar2007} were employed.
The resulting background levels for reactions with a neutron in the final state are listed in Table~\ref{table:NeutrinoCCBackgrounds}. We estimate $4.0 \pm 0.9$\ events in total.
The tagging efficiencies of the muon-decay-coincidence signature are calculated to be 78.6\% and $(77.5 \pm 0.2)\%$ from mean life times in carbon of positive and negative muons respectively. The residual background, including an untagged contribution from $(7.1 \pm 1.4)\%$ of negative muons which capture rather than decay, is $0.9 \pm 0.2$\ events. 
The observed rate of a muon decay signal is $\sim$0.2\ ${\rm events\,(kton\mathchar`-year)^{-1}}$, which is comparable to the backgrounds from ``invisible'' muons in Super-Kamiokande in this energy range~\cite{Malek2003}.

The most challenging background to estimate is that from the NC interactions of all neutrino species with carbon.
In these reactions the neutrino transfers only a fraction of its energy to the final products.
It can eject a neutron from the carbon nucleus, leaving it in an excited state with multiple decay modes. We used the following procedure to calculate the contribution from this background: we integrated the momentum transfer from a neutrino to a quasi-free neutron over the entire atmospheric neutrino spectra~\cite{Honda2007}
using cross sections from~\cite{Ahrens1987}. We then accounted for the neutron binding energies for P-shell (18.7\,MeV) and S-shell (41.7\,MeV) configurations and the corresponding shell populations. We also assumed that the neutron was removed from the carbon atom,
leaving it in an excited state.
All de-excitation modes reported in~\cite{Kamyshkov2003} were taken into account.
For each final product we converted the particle energy to visible energy in the detector using an energy scale model that includes nonlinearities from scintillator quenching.
Most of the outgoing neutrons have a kinetic energy less than 200\,MeV, and the resulting visible energies are concentrated in the lower energy region, typically less than 100\,MeV. 
The de-excitation of $^{11}{\rm C^{*}}$ is dominated by 2\,MeV gamma-ray emission, which has little effect on the energy spectral shape. In the analysis energy window, the position separation of energy depositions between prompt and delayed signals is $\sim60\,{\rm cm}$, which is larger than that for the thermal neutron case ($\sim40\,{\rm cm}$).
We calculate the contribution from this background to be 16.4 events with an estimated systematic uncertainty of 29\% which
is driven by uncertainties in the atmospheric neutrino flux and the cross section of NC neutrino interactions; see Table~\ref{table:NeutrinoNCBackgrounds}.

We also attempted to estimate the NC background using the NUANCE software tool (version 3), which simulates neutrino interactions and related processes~\cite{Casper2002}. However, we found the code overestimates this background rate by a factor of $\sim$2 relative to the above calculation in the energy region under study, mainly due to an inaccurate cross section for intra-nuclear nucleon re-scattering and an unexpected $\sim$25\,MeV positive offset of outgoing neutron energies. 
We therefore do not use the NUANCE-based estimation in this analysis.

\section{Data Interpretation}

\begin{figure}
\begin{center}
\includegraphics[angle=270,width=0.75\columnwidth]{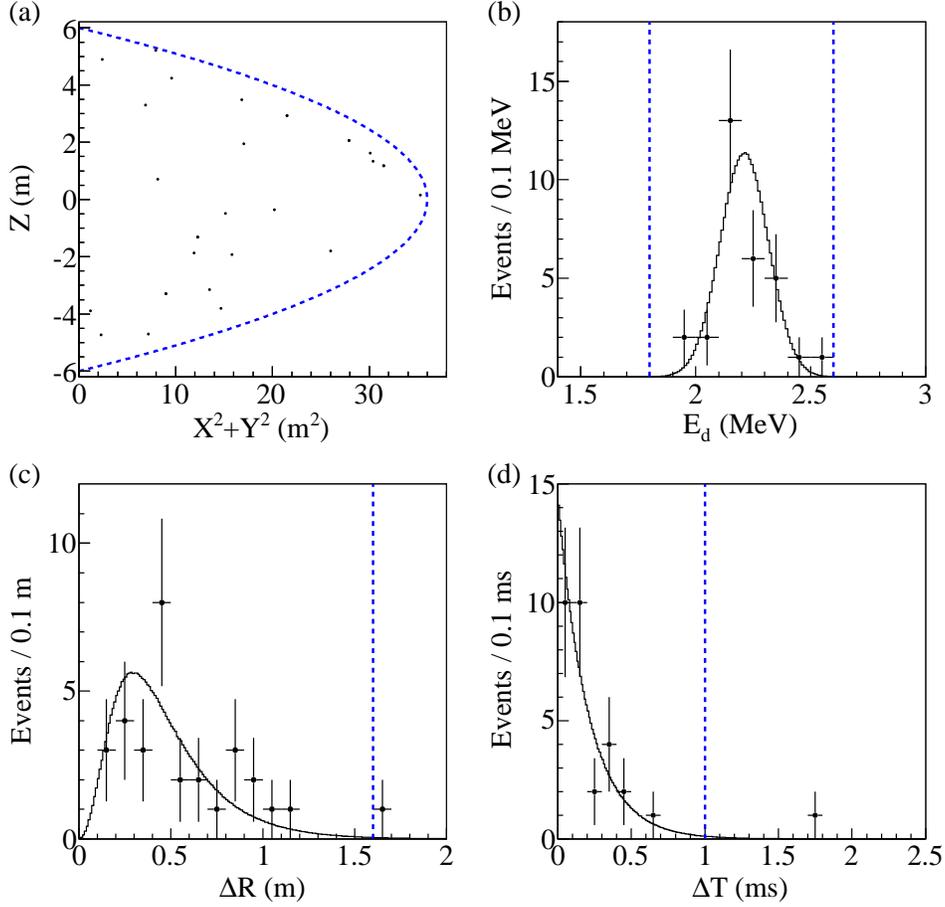}
\end{center}
\caption{Event distribution after all cuts: (a) prompt event position, (b) delayed energy, (c) spatial correlation, and (d) time correlation. The data are compared to the expected $\overline{\nu}_{e}$ signal generated by the Monte Carlo simulation (lines). Dashed lines indicate the selection criteria for $\overline{\nu}_{e}$ candidates.}
\label{figure:EventDistribution}
\end{figure}

\begin{figure}
\begin{center}
\includegraphics[angle=270,width=0.75\columnwidth]{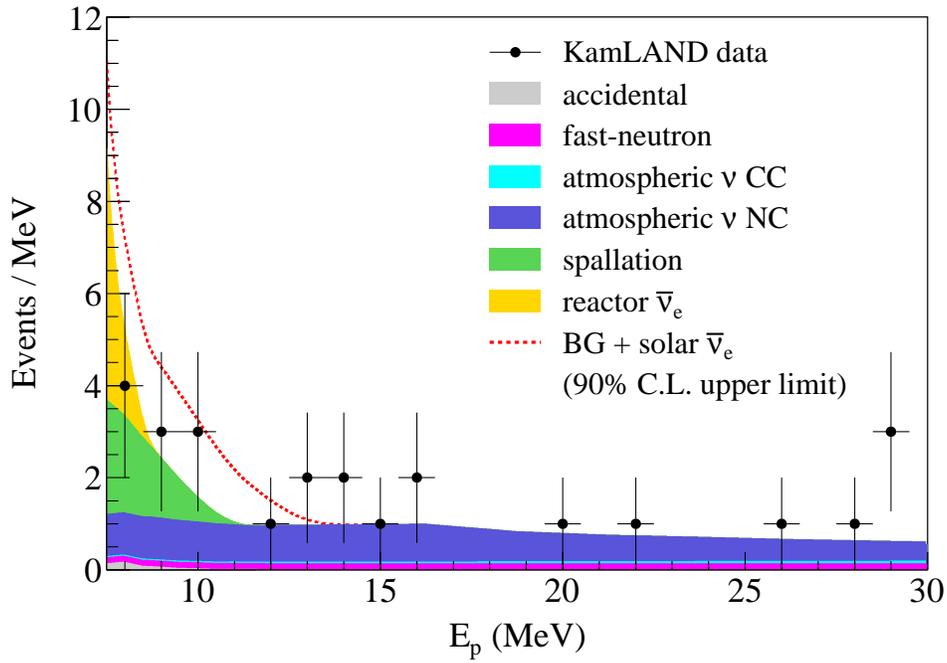}
\end{center}
\caption{Prompt energy distribution of selected $\overline{\nu}_{e}$ candidates in 1 MeV bins together with the best-fit backgrounds (filled areas) and 90\% CL upper limit of solar $\overline{\nu}_{e}$'s (red dashed line), which includes the detector response. The background histograms are cumulative and the solar $\overline{\nu}_{e}$ histogram sits on top of the background total.}
\label{figure:EnergyDistribution}
\end{figure}

We observe 25 events after the cuts described in Section~\ref{section:EventSelection}. Figure~\ref{figure:EventDistribution} illustrates the distribution for the prompt event position, the delayed energy, the spatial correlation, and the time correlation. There are five two-neutron candidates which may be caused by $^{12}{\rm C}(n, 2n)^{11}{\rm C}$ reaction of fast neutrons. The estimated number of backgrounds for $\overline{\nu}_{e}$ detection summarized in Table~\ref{table:BackgroundSummary} is $26.9 \pm 5.7$ events in the prompt energy window \mbox{$7.5\,{\rm MeV}<E_{\rm p}<30.0\,{\rm MeV}$}.  Figure~\ref{figure:EnergyDistribution} shows the event distribution as a function of prompt energy. The data set presented here contains 16 times more statistics than the first KamLAND publication on this subject, allowing us to verify the expected background contribution in the analyzed energy window.
The data are analyzed using an unbinned maximum likelihood fit to the event spectrum.
The estimate for $^{9}$Li and reactor $\overline{\nu}_{e}$ are rather robust, 
on the other hand,
reliable data for NC interactions in the energy range of interest do not exist and the method we used to calculate this background contribution has large uncertainties. 
To avoid possible bias from modeling in the NC background calculation, the normalization of the NC events is a free parameter in the spectral fits. 

\begin{figure}
\begin{center}
\includegraphics[angle=270,width=0.75\columnwidth]{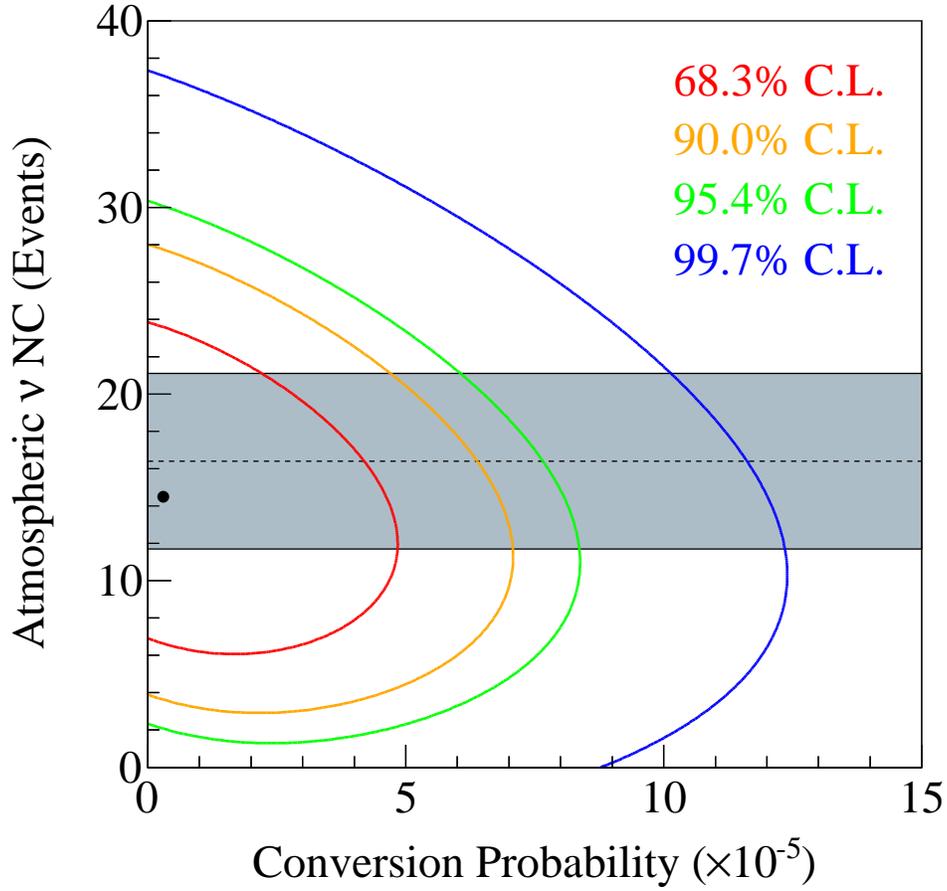}
\end{center}
\caption{Allowed region, with the best-fit point ($3.0 \times 10^{-6}$, $14.5$), for the unconstrained neutral current (NC) background and the probability of solar neutrino conversion from the KamLAND data in the energy range of \mbox{$8.3\,{\rm MeV} < E_{\overline{\nu}_{e}} < 31.8\,{\rm MeV}$}. The confidence level (CL) is shown for two degrees of freedom. The gray shaded region indicates the $\pm1\sigma$ prediction from the NC background calculation.}
\label{figure:ConversionProbability}
\end{figure}

From the unbinned maximum likelihood fit, the allowed region for the NC background and the conversion probability from $\nu_{e}$ to $\overline{\nu}_{e}$ 
is shown in Figure~\ref{figure:ConversionProbability}, assuming an unoscillated $^{8}{\rm B}$ neutrino flux of $5.94 \times 10^{6}~{\rm cm^{-2} s^{-1}}$~\cite{Pena-Garay2008}.
For the NC-floated normalization analysis, the upper limit for neutrino conversion is \mbox{$5.3\times10^{-5}$}\ at 90\% CL, which corresponds to a solar $\overline{\nu}_{e}$ flux of $93~{\rm cm^{-2} s^{-1}}$ or an event rate of $1.6\ {\rm events\,(kton\mathchar`-year)^{-1}}$ above the energy threshold (\mbox{$E_{\overline{\nu}_{e}} \ge 8.3~{\rm MeV}$}; containing 29.5\% of the total $^{8}{\rm B}$ neutrino flux). This limit is a factor 2.5 improvement over the previous limit in~\cite{Bellini2011} due to 24 times more exposure. For comparison, the rate analysis in the energy range \mbox{$8.3\,{\rm MeV} < E_{\overline{\nu}_{e}} < 15.0\,{\rm MeV}$} gives a slightly more stringent limit of $1.4~{\rm events\ (kton\mathchar`-year)^{-1}}$, if we use all the constraints on the background estimates including the NC background (Table~\ref{table:BackgroundSummary}). The fitted NC background assuming zero solar $\overline{\nu}_{e}$ events is $14.8^{+5.8}_{-5.4}$ events, which is in good agreement with the calculation ($16.4 \pm 4.7$ events).

The probability for solar neutrino conversion can be predicted by the models of spin flavor precession and Mikheyev-Smirnov-Wolfenstein large mixing angle solution oscillations in the Sun. 
If the conversion model for $^{8}{\rm B}$ neutrinos of Equation~(\ref{equation:FlavorPrecession}) 
without the distortion of the $^{8}{\rm B}$ spectrum 
is assumed, we obtain the following limit on the product of the neutrino magnetic moment $(\mu)$ and the transverse solar magnetic field in the region of neutrino production $(B_{T})$:
\begin{equation}
\label{equation:ProductLimit}
\frac{\mu}{10^{-12}\mu_{B}}
\frac{B_{T}(0.05R_{\odot})}{10\:\mbox{kG}}
< 5.9 \times 10^{2}, 
\end{equation}
using the value of $34^{\circ}$ for the mixing angle~\cite{Gando2011}.
The current best limit on the neutrino magnetic moment is from the GEMMA spectrometer, $\mu_{\overline{\nu}_{e}} < 3.2 \times 10^{-11}\,\mu_{B}$ at 90\% CL~\cite{Beda2010}. Lack of knowledge of the value of $B_{T}$ limits KamLAND sensitivity to the neutrino magnetic moment.

\begin{figure}
\begin{center}
\includegraphics[angle=270,width=0.75\columnwidth]{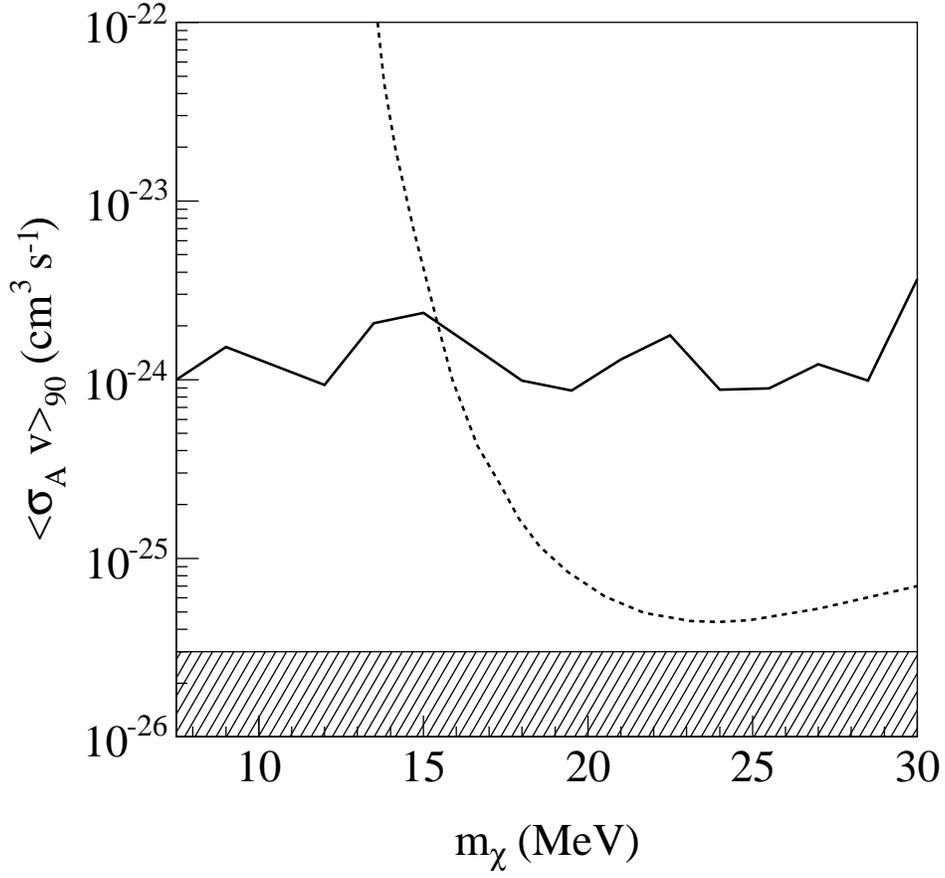}
\end{center}
\caption{Upper limits at 90\% CL on the dark matter annihilation cross section from KamLAND (solid line) and Super-Kamiokande (dashed line)~\cite{Palomares2008}. The shaded curve shows the natural scale of the annihilation cross section.}
\label{figure:DarkMatter}
\end{figure}

There data also test other potential $\overline{\nu}_{e}$ sources. 
Assuming an energy spectrum from the reference model~\cite{Ando2004}, which is consistent with a recent model reducing the cosmological uncertainties~\cite{Horiuchi2009}, we found an upper limit for the diffuse supernova $\overline{\nu}_{e}$ flux of $139~{\rm cm^{-2} s^{-1}}$ at 90\% CL in the analyzed energy range. 
This limit is weaker than our solar $\overline{\nu}_e$ flux limit 
due to the strong anticorrelation between the signal and NC background events amplified by the similarity in their spectral shapes. This flux limit corresponds to about 36 times the model prediction~\cite{Ando2004}, indicating poor statistical power in constraining the cosmological models using the current KamLAND data. The upper limit for the monochromatic $\overline{\nu}_{e}$ flux at each energy can be translated to a limit for the dark matter annihilation cross section~\cite{Palomares2008}. 
The dark matter annihilation limit varies weakly over the dark matter mass range due to limited statistics. We obtain \mbox{$\left< \sigma_{A} v \right> < $ (1--3) $\times 10^{-24}\,\rm{cm}^{3} \rm{s}^{-1}$} at 90\% CL in the mass range $8.3~{\rm MeV} < m_{\chi} < 31.8~{\rm MeV}$, as shown in Figure~\ref{figure:DarkMatter}.
This is the most stringent constraint on the annihilation cross section below 15 MeV.

Finally, we also present model-independent upper limits for $\overline{\nu}_{e}$ fluxes, as shown in Figure~\ref{figure:UpperLimits}. The limits are given at 90\% CL based on the rate analysis using the Feldman-Cousins approach~\cite{Feldman1998} with 1\,MeV energy bins, including all the constraints on the background estimates in Table~\ref{table:BackgroundSummary}. The KamLAND data provide the best limits in the presented energy range \mbox{$8.3\,{\rm MeV} < E_{\overline{\nu}_{e}} < 18.3\,{\rm MeV}$}, owing to the efficient $\overline{\nu}_{e}$ detection by the DC method and large exposure. Given that data are background limited, mainly from the atmospheric neutrino NC interactions, accumulation of additional statistics is unlikely to improve this limit significantly. 

In conclusion,
we report the spectrum of high-energy $\overline{\nu}_{e}$ candidates found in the KamLAND data set accumulated over more than eight years of detector operation.
The live time exposure corresponds to 4.53\,kton-year.
In the energy range from 8.3~MeV to 31.8~MeV, no excess of $\overline{\nu}_{e}$ events over the expected background consisting of mostly 
atmospheric neutrino NC interactions, cosmogenically induced radioactivity, and reactor neutrinos were detected.
The data allow significantly improved limits on solar $\overline{\nu}_{e}$ conversion probability, and on DSNF and annihilation cross section of dark matter below 15 MeV.
The present level of background indicates limitations for future studies of $\overline{\nu}_{e}$'s in this energy range using KamLAND.

While a better detector location could eliminate $^{9}$Li background and suppress reactor neutrino background, atmospheric neutrino NC interactions will continue to present significant challenges for next-generation large LS detectors. In water Cherenkov detectors, like Super-Kamiokande, the contribution from NC backgrounds is expected to be small, because the recoil protons by knocked-out neutrons from $^{16}{\rm O}$ should be below the Cherenkov energy threshold. If gadolinium is added, Super-Kamiokande will overcome the problem of large backgrounds from solar neutrinos, spallation products, and invisible muon decays, by the DC technique~\cite{Beacom2004}, and gain the ability to detect the diffuse supernova neutrino signals. A possible future LENA experiment~\cite{Wurm2007}, which will have a detector with about 50 kton of LS,  also aims to measure DSNF, reducing NC backgrounds by the $^{11}{\rm C}$ tagging methods~\cite{Wurm2011}. For success, the detector design needs to be optimized to maximize the efficiency of the NC background rejection.

\acknowledgments

The KamLAND experiment is supported by the Grant-in-Aid for Specially Promoted Research under grant no. 16002002 of the Japanese Ministry of Education, Culture, Sports, Science and Technology; the World Premier International Research Center Initiative (WPI Initiative), MEXT, Japan; and under the U.S. Department of Energy (DOE) grants no. DE-FG03-00ER41138, DE-AC02-05CH11231, and DE-FG02-01ER41166, as well as other DOE grants to individual institutions. The reactor data are provided by courtesy of the following electric associations in Japan: Hokkaido, Tohoku, Tokyo, Hokuriku, Chubu, Kansai, Chugoku, Shikoku, and Kyushu Electric Power Companies, Japan Atomic Power Company, and Japan Atomic Energy Agency. 
The Kamioka Mining and Smelting Company has provided service for activities in the mine.

\section*{APPENDIX}
\begin{figure}
\begin{center}
\includegraphics[angle=270,width=0.75\columnwidth]{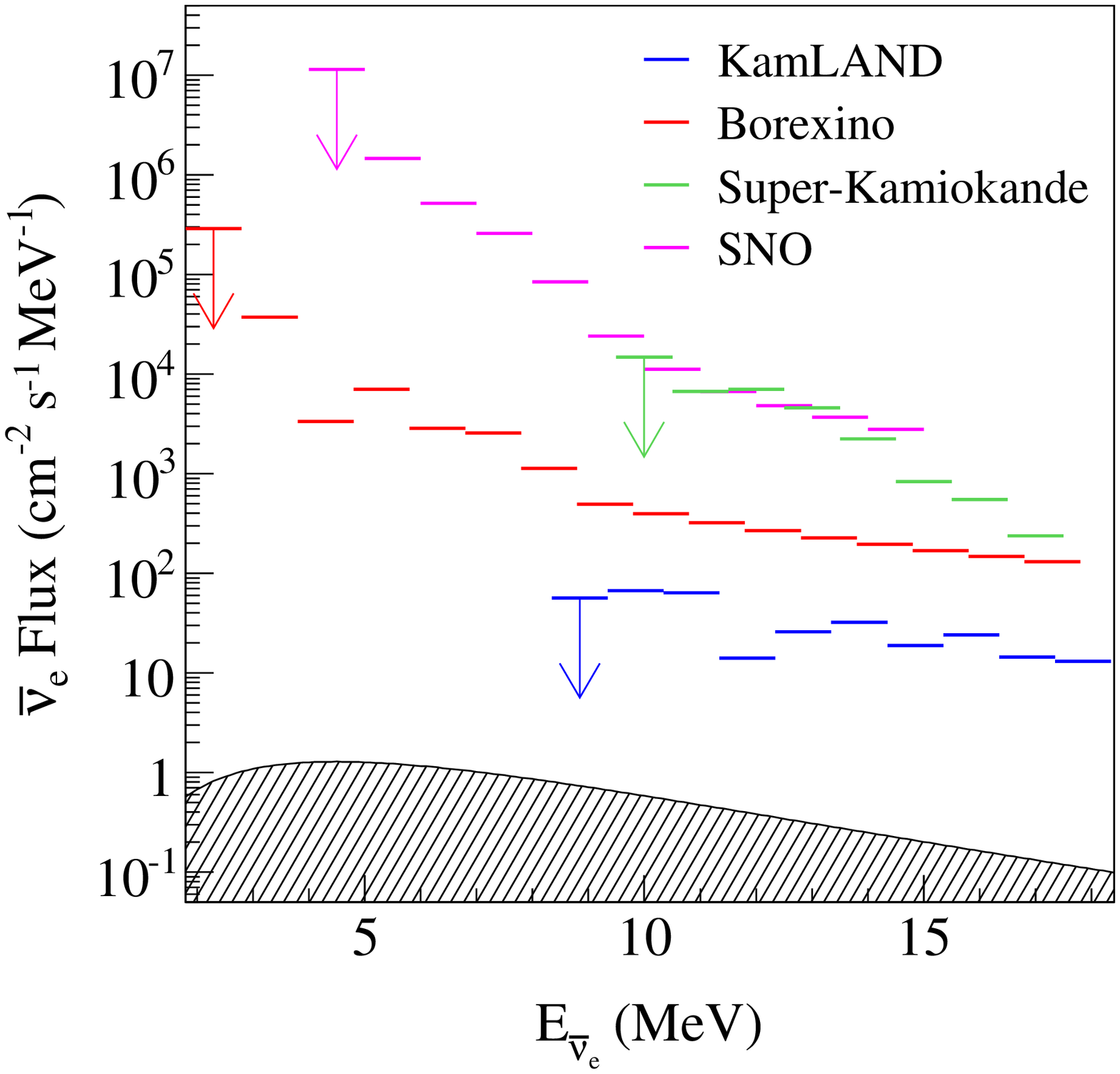}
\end{center}
\caption{Model independent upper limits at 90\% CL on the $\overline{\nu}_{e}$ flux from KamLAND (blue line), Borexino~\cite{Bellini2011} (red line), SNO~\cite{Aharmim2004} (magenta line), and Super-Kamiokande~\cite{Gando2003} (green line). The $\overline{\nu}_{e}$ flux limit depends on the event rate limit and the cross section of $\overline{\nu}_{e}\mathchar`-p$. The shaded curve shows the diffuse supernova $\overline{\nu}_{e}$ flux for the reference model prediction~\cite{Ando2004}. Above the shown energy range, the Super-Kamiokande data give more stringent upper limit of $1.2\,{\rm cm}^{-2} {\rm s}^{-1}$ for the diffuse supernova $\overline{\nu}_{e}$ flux, owing to less amount of backgrounds from muon spallation products~\cite{Malek2003}.}
\label{figure:UpperLimits}
\end{figure}

The model-independent upper limits for $\overline{\nu}_{e}$ fluxes provided for each energy may be useful to give an estimate of upper limits for various $\overline{\nu}_{e}$ sources. For example, one can easily test one's own model with a certain energy spectrum by an appropriate data integration. Table~\ref{table:UpperLimits} lists the 1 MeV binned upper limits shown in Figure~\ref{figure:UpperLimits}. The binned $\chi^{2}$ is defined as 
 \begin{eqnarray}
 \chi^{2} & = & \sum_{i} \frac{\nu_{i}^{2}}{(u_{i} / \sqrt{2.71})^{2}}
\label{equation:chi2}
 \end{eqnarray}
 where $\nu_{i}$ is the model expectation for each energy bin, $u_{i}$ is the KamLAND upper limit at 90\% CL, and $\sqrt{2.71}$ is the conversion factor of limits from 90\% CL to 1$\sigma$ CL. 
This binned $\chi^{2}$ analysis give the upper limit of the solar $\overline{\nu}_{e}$ flux $(< 87~{\rm cm^{-2} s^{-1}})$, and approximately reproduces the limit which is based on the unbinned maximum likelihood method including all background and systematic uncertainties. On the other hand, a limit for the diffuse supernova $\overline{\nu}_{e}$ flux based on Equation~(\ref{equation:chi2}) will be optimistic, because Table~\ref{table:UpperLimits} data include the constraints on the NC background estimate.

\bibliography{HighEnergyAntineutrino}

\begin{thebibliography}{39}
\expandafter\ifx\csname natexlab\endcsname\relax\def\natexlab#1{#1}\fi

\bibitem[{Abe {et~al.}(2010)Abe, Enomoto, Furuno, Gando, Ikeda, Inoue, Kibe,
  Kishimoto, Koga, Minekawa, Mitsui, Nakajima, Nakajima, Nakamura, Nakamura,
  Shimizu, Shimizu, Shirai, Suekane, Suzuki, Takemoto, Tamae, Terashima,
  Watanabe, Yonezawa, Yoshida, \& Kozlov}]{Abe2010}
Abe, S., {et~al.} 2010, Phys. Rev. C, 81, 025807

\bibitem[{Aharmim {et~al.}(2004)Aharmim, Ahmed, Beier, Bellerive, Biller,
  Boger, Boulay, Bowles, Brice, Bullard, Chan, Chen, Chen, Cleveland, Cox, Dai,
  Dalnoki-Veress, Doe, Dosanjh, Doucas, Dragowsky, Duba, Duncan, Dunford,
  Dunmore, Earle, Elliott, Evans, Ewan, Farine, Fergani, Fleurot, Formaggio,
  Fowler, Frame, Frati, Fulsom, Gagnon, Graham, Grant, Hahn, Hallin, Hallman,
  Hamer, Handler, Hargrove, Harvey, Hazama, Heeger, Heintzelman, Heise, Helmer,
  Hemingway, Hime, Howe, Jagam, Jelley, Klein, Kormos, Kos, Kr\"uger, Krauss,
  Krumins, Kutter, Kyba, Labranche, Lange, Law, Lawson, Lesko, Leslie, Levine,
  Luoma, MacLellan, Majerus, Mak, Maneira, Marino, McCauley, McDonald, McGee,
  McGregor, Mifflin, Miknaitis, Miller, Moffat, Nally, Neubauer, Nickel, Noble,
  Norman, Oblath, Okada, Ollerhead, Orrell, Oser, Ouellet, Peeters, Poon,
  Rielage, Robertson, Robertson, Rollin, Rosendahl, Rusu, Schwendener, Simard,
  Simpson, Sims, Sinclair, Skensved, Smith, Starinsky, Stokstad, Stonehill,
  Tafirout, Takeuchi, Te\ifmmode \check{s}\else
  \v{s}\fi{}i\ifmmode~\acute{c}\else \'{c}\fi{}, Thomson, Tsui, Van~Berg,
  Van~de Water, Virtue, Wall, Waller, Waltham, Tseung, Wark, West, Wilhelmy,
  Wilkerson, Wilson, Wittich, Wouters, Yeh, \& Zuber}]{Aharmim2004}
Aharmim, B., {et~al.} 2004, Phys. Rev. D, 70, 093014

\bibitem[{Aharmim {et~al.}(2006)Aharmim, Ahmed, Anthony, Beier, Bellerive,
  Bergevin, Biller, Boulay, Chan, Chen, Chen, Cleveland, Cox, Currat, Dai,
  Dalnoki-Veress, Deng, Detwiler, DiMarco, Doe, Doucas, Drouin, Duncan,
  Dunford, Dunmore, Earle, Evans, Ewan, Farine, Fergani, Fleurot, Ford,
  Formaggio, Gagnon, Goon, Graham, Guillian, Hahn, Hallin, Hallman, Harvey,
  Hazama, Heeger, Heintzelman, Heise, Helmer, Hemingway, Henning, Hime, Howard,
  Howe, Huang, Jagam, Jelley, Klein, Kormos, Kos, Kr{\"u}ger, Kraus, Krauss,
  Kutter, Kyba, Labranche, Lange, Law, Lawson, Lesko, Leslie, Loach, Luoma,
  MacLellan, Majerus, Mak, Maneira, Marino, Martin, McCauley, McDonald, McGee,
  Mifflin, Miknaitis, Miller, Monreal, Nickel, Noble, Norman, Oblath, Okada,
  O'Keeffe, Gann, Oser, Ott, Peeters, Poon, Prior, Rielage, Robertson,
  Robertson, Rollin, Schwendener, Secrest, Seibert, Simard, Sims, Sinclair,
  Skensved, Stokstad, Stonehill, Te{\v s}i{\'c}, Tolich, Tsui, Berg, de~Water,
  VanDevender, Virtue, Walker, Wall, Waller, Tseung, Wark, Wendland, West,
  Wilkerson, Wilson, Wouters, Wright, Yeh, Zhang, \& Zuber}]{Aharmim2006}
Aharmim, B., {et~al.} 2006, ApJ, 653, 1545

\bibitem[{Ahrens {et~al.}(1987)Ahrens, Aronson, Connolly, Gibbard, Murtagh,
  Murtagh, Terada, White, Callas, Cutts, Hoftun, Diwan, Lanou, Shinkawa,
  Kurihara, Amako, Kabe, Nagashima, Suzuki, Tatsumi, Yamaguchi, Abe, Beier,
  Doughty, Durkin, Heagy, \& Hurley}]{Ahrens1987}
Ahrens, L.~A., {et~al.} 1987, Phys. Rev. D, 35, 785

\bibitem[{Akhmedov \& Pulido(2003)}]{Akhmedov2003}
Akhmedov, E.~K., \& Pulido, J. 2003, Phys. Lett. B, 553, 7

\bibitem[{Ando \& Sato(2004)}]{Ando2004}
Ando, S., \& Sato, K. 2004, New J. Phys., 6, 170

\bibitem[{Araki {et~al.}(2005)Araki, Enomoto, Furuno, Gando, Ichimura, Ikeda,
  Inoue, Kishimoto, Koga, Koseki, Maeda, Mitsui, Motoki, Nakajima, Ogawa,
  Ogawa, Owada, Ricol, Shimizu, Shirai, Suekane, Suzuki, Tada, Takeuchi, Tamae,
  Tsuda, Watanabe, Busenitz, Classen, Djurcic, Keefer, Leonard, Piepke,
  Yakushev, Berger, Chan, Decowski, Dwyer, Freedman, Fujikawa, Goldman, Gray,
  Heeger, Hsu, Lesko, Luk, Murayama, O'Donnell, Poon, Steiner, Winslow, Mauger,
  McKeown, Vogel, Lane, Miletic, Guillian, Learned, Maricic, Matsuno, Pakvasa,
  Horton-Smith, Dazeley, Hatakeyama, Rojas, Svoboda, Dieterle, Detwiler,
  Gratta, Ishii, Tolich, Uchida, Batygov, Bugg, Efremenko, Kamyshkov, Kozlov,
  Nakamura, Karwowski, Markoff, Nakamura, Rohm, Tornow, Wendell, Chen, Wang, \&
  Piquemal}]{Araki2005b}
Araki, T., {et~al.} 2005, Nature, 436, 499

\bibitem[{Athar {et~al.}(2007)Athar, Ahmad, \& Singh}]{Athar2007}
Athar, M.~S., Ahmad, S., \& Singh, S.~K. 2007, Phys. Rev. D, 75, 093003

\bibitem[{Beacom(2010)}]{Beacom2010}
Beacom, J.~F. 2010, Annu. Rev. Nucl. Part. Sci., 60, 439

\bibitem[{Beacom \& Bell(2002)}]{Beacom2002}
Beacom, J.~F., \& Bell, N.~F. 2002, Phys. Rev. D, 65, 113009

\bibitem[{Beacom \& Vagins(2004)}]{Beacom2004}
Beacom, J.~F., \& Vagins, M.~R. 2004, Phys. Rev. Lett., 93, 171101

\bibitem[{Beda {et~al.}(2010)Beda, Brudanin, Egorov, Medvedev, Pogosov,
  Shirchenko, \& Starostin}]{Beda2010}
Beda, A.~G., Brudanin, V.~B., Egorov, V.~G., Medvedev, D.~V., Pogosov, V.~S.,
  Shirchenko, M.~V., \& Starostin, A.~S. 2010, arXiv:1005.2736

\bibitem[{Bellini {et~al.}(2010)Bellini, Benziger, Bonetti, Avanzini,
  Caccianiga, Cadonati, Calaprice, Carraro, Chavarria, Dalnoki-Veress,
  D'Angelo, Davini, de~Kerret, Derbin, Etenko, Fiorentini, Fomenko, Franco,
  Galbiati, Gazzana, Ghiano, Giammarchi, Goeger-Neff, Goretti, Guardincerri,
  Hardy, Ianni, Ianni, Joyce, Kobychev, Koshio, Korga, Kryn, Laubenstein,
  Leung, Lewke, Litvinovich, Loer, Lombardi, Ludhova, Machulin, Manecki,
  Maneschg, Manuzio, Meindl, Meroni, Miramonti, Misiaszek, Montanari, Muratova,
  Oberauer, Obolensky, Ortica, Pallavicini, Papp, Perasso, Perasso, Pocar,
  Raghavan, Ranucci, Razeto, Re, Ricci, Risso, Romani, Rountree, Sabelnikov,
  Saldanha, Salvo, Schert, Simgen, Skorokhvatov, Smirnov, Sotnikov,
  Sukhotin, Suvorov, Tartaglia, Testera, Vignaud, Vogelaar, von Feilitzsch,
  Winter, Wojcik, Wright, Wurm, Xu, Zaimidoroga, Zavatarelli, \&
  Zuzel}]{Bellini2010}
Bellini, G., {et~al.} 2010, Phys. Lett. B, 687, 299

\bibitem[{Bellini {et~al.}(2011)Bellini, Benziger, Bonetti, Buizza~Avanzini,
  Caccianiga, Cadonati, Calaprice, Carraro, Chavarria, Chepurnov, D'Angelo,
  Davini, Derbin, Etenko, Fomenko, Franco, Galbiati, Gazzana, Ghiano,
  Giammarchi, Goeger-Neff, Goretti, Guardincerri, Hardy, Ianni, Ianni, Joyce,
  Kobychev, Korablev, Koshio, Korga, Kryn, Laubenstein, Lewke, Litvinovich,
  Loer, Lombardi, Ludhova, Machulin, Manecki, Maneschg, Manuzio, Meindl,
  Meroni, Miramonti, Misiaszek, Montanari, Muratova, Oberauer, Obolensky,
  Ortica, Pallavicini, Papp, Perasso, Perasso, Pocar, Raghavan, Ranucci,
  Razeto, Re, Risso, Romani, Rountree, Sabelnikov, Saldanha, Salvo, Schert,
  Simgen, Skorokhvatov, Smirnov, Sotnikov, Sukhotin, Suvorov, Tartaglia,
  Testera, Vignaud, Vogelaar, von Feilitzsch, Winter, Wojcik, Wright, Wurm, Xu,
  Zaimidoroga, Zavatarelli, \& Zuzel}]{Bellini2011}
Bellini, G., {et~al.} 2011, Phys. Lett. B, 696, 191

\bibitem[{Casper(2002)}]{Casper2002}
Casper, D. 2002, Nucl. Phys. B Proc. Suppl., 112, 161

\bibitem[{Couvidat {et~al.}(2003)Couvidat, Turck-Chi{\`e}ze, \&
  Kosovichev}]{Couvidat2003}
Couvidat, S., Turck-Chi{\`e}ze, S., \& Kosovichev, A.~G. 2003, ApJ, 599, 1434

\bibitem[{Cowan {et~al.}(1956)Cowan, Reines, Harrison, Kruse, \&
  McGuire}]{Cowan1956}
Cowan, C.~L., J., Reines, F., Harrison, F.~B., Kruse, H.~W., \& McGuire, A.~D.
  1956, Science, 124, 103

\bibitem[{Feldman \& Cousins(1998)}]{Feldman1998}
Feldman, G.~J., \& Cousins, R.~D. 1998, Phys. Rev. D, 57, 3873

\bibitem[{Gando {et~al.}(2011)Gando, Gando, Ichimura, Ikeda, Inoue, Kibe,
  Kishimoto, Koga, Minekawa, Mitsui, Morikawa, Nagai, Nakajima, Nakamura,
  Narita, Shimizu, Shimizu, Shirai, Suekane, Suzuki, Takahashi, Takahashi,
  Takemoto, Tamae, Watanabe, Xu, Yabumoto, Yoshida, Yoshida, Enomoto, Kozlov,
  Murayama, Grant, Keefer, Piepke, Banks, Bloxham, Detwiler, Freedman,
  Fujikawa, Han, Kadel, O'Donnell, Steiner, Dwyer, McKeown, Zhang, Berger,
  Lane, Maricic, Miletic, Batygov, Learned, Matsuno, Sakai, Horton-Smith,
  Downum, Gratta, Efremenko, Perevozchikov, Karwowski, Markoff, Tornow, Heeger,
  \& Decowski}]{Gando2011}
Gando, A., {et~al.} 2011, Phys. Rev. D, 83, 052002

\bibitem[{Gando {et~al.}(2003)Gando, Fukuda, Fukuda, Ishitsuka, Itow, Kajita,
  Kameda, Kaneyuki, Kobayashi, Koshio, Miura, Moriyama, Nakahata, Nakayama,
  Namba, Obayashi, Okada, Ooyabu, Saji, Sakurai, Shiozawa, Suzuki, Takeuchi,
  Takeuchi, Totsuka, Yamada, Desai, Earl, Kearns, Messier, Stone, Sulak,
  Walter, Goldhaber, Barszczak, Casper, Gajewski, Kropp, Mine, Liu, Smy, Sobel,
  Vagins, Gago, Ganezer, Hill, Keig, Ellsworth, Tasaka, Kibayashi, Learned,
  Matsuno, Takemori, Hayato, Ichikawa, Ishii, Kobayashi, Maruyama, Nakamura,
  Oyama, Sakuda, Yoshida, Kohama, Iwashita, Suzuki, Inagaki, Kato, Nakaya,
  Nishikawa, Haines, Dazeley, Hatakeyama, Svoboda, Blaufuss, Chen, Goodman,
  Guillian, Sullivan, Turcan, Scholberg, Habig, Ackermann, Jung, Martens,
  Malek, Mauger, McGrew, Sharkey, Viren, Yanagisawa, Toshito, Mitsuda, Miyano,
  Shibata, Kajiyama, Nagashima, Nitta, Takita, Kim, Kim, Yoo, Okazawa,
  Ishizuka, Etoh, Hasegawa, Inoue, Ishihara, Shirai, Suzuki, Koshiba,
  Hatakeyama, Ichikawa, Koike, Nishijima, Ishino, Morii, Nishimura, Watanabe,
  Kielczewska, Berns, Boyd, Stachyra, \& Wilkes}]{Gando2003}
Gando, Y., {et~al.} 2003, Phys. Rev. Lett., 90, 171302

\bibitem[{Hahn {et~al.}(1989)Hahn, Schreckenbach, Gelletly, von Feilitzsch,
  Colvin, \& Krusche}]{Hahn1989}
Hahn, A.~A., Schreckenbach, K., Gelletly, W., von Feilitzsch, F., Colvin, G.,
  \& Krusche, B. 1989, Phys. Lett. B, 218, 365

\bibitem[{Honda {et~al.}(2007)Honda, Kajita, Kasahara, Midorikawa, \&
  Sanuki}]{Honda2007}
Honda, M., Kajita, T., Kasahara, K., Midorikawa, S., \& Sanuki, T. 2007, Phys.
  Rev. D, 75, 043006

\bibitem[{Horiuchi {et~al.}(2009)Horiuchi, Beacom, \& Dwek}]{Horiuchi2009}
Horiuchi, S., Beacom, J.~F., \& Dwek, E. 2009, Phys. Rev. D, 79, 083013

\bibitem[{Kamyshkov \& Kolbe(2003)}]{Kamyshkov2003}
Kamyshkov, Y., \& Kolbe, E. 2003, Phys. Rev. D, 67, 076007

\bibitem[{Kim \& Cheoun(2009)}]{Kim2009}
Kim, K.~S., \& Cheoun, M.-K. 2009, Phys. Lett. B, 679, 330

\bibitem[{Kume {et~al.}(1983)Kume, Sawaki, Ito, Arisaka, Kajita, Nishimura, \&
  Suzuki}]{Kume1983}
Kume, H., Sawaki, S., Ito, M., Arisaka, K., Kajita, T., Nishimura, A., \&
  Suzuki, A. 1983, Nucl. Instr. Meth. Phys. Res., 205, 443

\bibitem[{Mack {et~al.}(2008)Mack, Jacques, Beacom, Bell, \&
  Y{\"u}ksel}]{Mack2008}
Mack, G.~D., Jacques, T.~D., Beacom, J.~F., Bell, N.~F., \& Y{\"u}ksel, H.
  2008, Phys. Rev. D, 78, 063542

\bibitem[{Malaney {et~al.}(1990)Malaney, Meyer, \& Butler}]{Malaney1990}
Malaney, R.~A., Meyer, B.~S., \& Butler, M.~N. 1990, ApJ, 352, 767

\bibitem[{Malek {et~al.}(2003)Malek, Morii, Fukuda, Fukuda, Ishitsuka, Itow,
  Kajita, Kameda, Kaneyuki, Kobayashi, Koshio, Miura, Moriyama, Nakahata,
  Nakayama, Namba, Okada, Ooyabu, Saji, Sakurai, Shiozawa, Suzuki, Takeuchi,
  Takeuchi, Totsuka, Yamada, \& Desai}]{Malek2003}
Malek, M., {et~al.} 2003, Phys. Rev. Lett., 90, 061101

\bibitem[{Okun {et~al.}(1986)Okun, Voloshin, \& Vysotsky}]{Okun1986}
Okun, L.~B., Voloshin, M.~B., \& Vysotsky, M.~I. 1986, Sov. Phys. JETP, 64, 446

\bibitem[{Palomares-Ruiz \& Pascoli(2008)}]{Palomares2008}
Palomares-Ruiz, S., \& Pascoli, S. 2008, Phys. Rev. D, 77, 025025

\bibitem[{Pena-Garay \& Serenelli(2008)}]{Pena-Garay2008}
Pena-Garay, C., \& Serenelli, A. 2008, arXiv:0811.2424

\bibitem[{Schreckenbach {et~al.}(1985)Schreckenbach, Colvin, Gelletly, \&
  Von~Feilitzsch}]{Schreckenbach1985}
Schreckenbach, K., Colvin, G., Gelletly, W., \& Von~Feilitzsch, F. 1985, Phys.
  Lett. B, 160, 325

\bibitem[{Serenelli(2010)}]{Serenelli2010}
Serenelli, A. 2010, Astrophys. Space Sci., 328, 13

\bibitem[{Totani \& Sato(1995)}]{Totani1995}
Totani, T., \& Sato, K. 1995, Astropart. Phys., 3, 367

\bibitem[{Vogel \& Beacom(1999)}]{Vogel1999}
Vogel, P., \& Beacom, J.~F. 1999, Phys. Rev. D, 60, 053003

\bibitem[{Vogel {et~al.}(1981)Vogel, Schenter, Mann, \& Schenter}]{Vogel1981}
Vogel, P., Schenter, G.~K., Mann, F.~M., \& Schenter, R.~E. 1981, Phys. Rev. C,
  24, 1543

\bibitem[{Wurm {et~al.}(2007)Wurm, von Feilitzsch, G{\"o}ger-Neff, Hochmuth,
  Undagoitia, Oberauer, \& Potzel}]{Wurm2007}
Wurm, M., von Feilitzsch, F., G{\"o}ger-Neff, M., Hochmuth, K.~A., Undagoitia,
  T.~M., Oberauer, L., \& Potzel, W. 2007, Phys. Rev. D, 75, 023007

\bibitem[{Wurm {et~al.}(2011)Wurm, Beacom, Bezrukov, Bick, Bl{\"u}mer, Choubey,
  Ciemniak, D'Angelo, Dasgupta, Dighe, Domogatsky, Dye, Eliseev, Enqvist,
  Erykalov, von Feilitzsch, Fiorentini, Fischer, G{\"o}ger-Neff, Grabmayr,
  Hagner, Hellgartner, Hissa, Horiuchi, Janka, Jaupart, Jochum, Kalliokoski,
  Kuusiniemi, Lachenmaier, Lazanu, Learned, Lewke, Lombardi, Lorenz,
  Lubsandorzhiev, Ludhova, Loo, Maalampi, Mantovani, Marafini, Maricic,
  Undagoitia, McDonough, Miramonti, \& Mirizzi}]{Wurm2011}
Wurm, M., {et~al.} 2011, arXiv:1104.5620

\end{thebibliography}

\clearpage

\clearpage

\clearpage

\begin{deluxetable}{ccc}
\tablecaption{Calculated Backgrounds for Atmospheric Neutrino CC Interactions with Prompt Energy between 7.5 MeV and 30.0\,MeV
\label{table:NeutrinoCCBackgrounds}
}
\tablewidth{0pt}
\tablehead{
\colhead{Reaction} & \colhead{Number of Events} & \colhead{Number of Untagged Events}
}
\startdata
$\overline{\nu}_{\mu}+p\rightarrow\mu^{+}+n$ & 2.1 & 0.5 \\
$\overline{\nu}_{\mu}+^{12}$C$\rightarrow\mu^{+}+n+^{11}$B & 0.7 & 0.2 \\
$\nu_{\mu}+^{12}$C$\rightarrow\mu^{-}+n+^{11}$N & 0.4 & 0.1\\
$\overline{\nu}_{\mu}+^{12}$C$\rightarrow\mu^{+}+n+^{11}$B$+\gamma$ & 0.4 & 0.08 \\
$\overline{\nu}_{\mu}+^{12}$C$\rightarrow\mu^{+}+n+^{7}$Li$+\alpha$ & 0.4 & 0.08 \\
$\overline{\nu}_{\mu}+^{12}$C$\rightarrow\mu^{+}+2n+^{10}$B & 0.02 & 0.005 \\
\hline
Total & $4.0\pm0.9$ & $0.9\pm0.2$ \\
\enddata
\\
\vspace{0.5 cm}
{\bf Note.} The numbers in the third column include the inefficiencies of the muon decay.
\end{deluxetable}

\begin{deluxetable}{ccc}
\tablecaption{Calculated Backgrounds for Atmospheric Neutrino NC Interactions with Prompt Energy between 7.5 MeV and 30.0\,MeV
\label{table:NeutrinoNCBackgrounds}
}
\tablewidth{0pt}
\tablehead{
\colhead{Reaction} & \colhead{Number of Events}
}
\startdata
$\nu(\overline{\nu})+^{12}$C$\rightarrow\nu(\overline{\nu})+n+^{11}$C$+\gamma$ & 13.2 \\
$\nu(\overline{\nu})+^{12}$C$\rightarrow\nu(\overline{\nu})+n+^{10}$B$+p$ & 1.4 \\
$\nu(\overline{\nu})+^{12}$C$\rightarrow\nu(\overline{\nu})+n+^{6}$Li$+\alpha+p$ & 1.4 \\
$\nu(\overline{\nu})+^{12}$C$\rightarrow\nu(\overline{\nu})+n+^{9}$Be$+2p$ & 0.3 \\
$\nu(\overline{\nu})+^{12}$C$\rightarrow\nu(\overline{\nu})+2n+^{10}$C$$ & 0.1 \\
\hline
Total & $16.4\pm4.7$ \\
\enddata
\\
\vspace{0.5 cm}
{\bf Note.} There is no muon in the final state, so the muon decay tagging is not useful unlike CC interactions.
\end{deluxetable}

\begin{deluxetable}{cc}
\tablecaption{Summary of the Estimated Backgrounds with Prompt Energy between 7.5 MeV and 30.0\,MeV
\label{table:BackgroundSummary}
}
\tablewidth{0pt}
\tablehead{
\colhead{Background} & \colhead{Number of Events} 
}
\startdata
Random coincidences & $0.22\pm0.01$ \\
Reactor $\overline{\nu}_{e}$ & $2.2\pm0.7$ \\
$^{9}$Li & $4.0\pm0.3$ \\
Fast neutron & $3.2\pm3.2$ \\
Atmospheric $\nu$\ (CC) & $0.9\pm0.2$ \\
Atmospheric $\nu$\ (NC) & $16.4\pm4.7$ \\
\hline
Total & $26.9\pm5.7$ \\
\enddata
\end{deluxetable}

\begin{deluxetable}{cc}
\tablecaption{Model Independent Upper Limit on the $\overline{\nu}_{e}$ Flux for Each Energy Bin from KamLAND, as Shown in Figure~\ref{figure:UpperLimits}
\label{table:UpperLimits}
}
\tablewidth{0pt}
\tablehead{
\colhead{Energy Range (MeV)} & \colhead{Upper Limit at 90\% CL (${{\rm cm}^{-2}{\rm s}^{-1}}$)} 
}
\startdata
8.3--9.3 & 56.2\\
9.3--10.3 & 67.1\\
10.3--11.3 & 63.8\\
11.3--12.3 & 14.0\\
12.3--13.3 & 25.8\\
13.3--14.3 & 32.2\\
14.3--15.3 & 18.9\\
15.3--16.3 & 24.1\\
16.3--17.3 & 14.5\\
17.3--18.3 & 13.0\\
\enddata
\end{deluxetable}

\end{document}